\documentclass{iopart}
\usepackage{graphicx,bm,epsfig}
\begin{document}
\title{Percolation and critical O($n$) loop configurations}
\author{Chengxiang Ding$^{1}$, 
Youjin Deng$^2$\footnotemark[1],
Wenan Guo$^{1}$\footnote[1]
{Corresponding authors: waguo@bnu.edu.cn, yjdeng@ustc.edu.cn}, 
and Henk W.J. Bl\"ote~$^{3}$} 
\address{$^{1}$Physics Department, Beijing Normal University,
Beijing 100875, P. R. China }
\address{$^{2}$Hefei National Laboratory for Physical 
Sciences at Microscale, 
Department of Modern Physics, University of Science and 
Technology of China, Hefei 230027, China }
\address{$^{3}$ Instituut Lorentz, Leiden University,
  P.O. Box 9506, 2300 RA Leiden, The Netherlands}

\date{\today} 
\begin{abstract}
We study a percolation problem based on critical loop configurations
of the  O($n$) loop model on the honeycomb lattice.
We define dual clusters as groups of sites on the dual triangular
lattice that are not separated by a loop, and investigate the 
the bond-percolation properties of these dual clusters.
The universal properties at the percolation threshold are argued to
match those of Kasteleyn-Fortuin random clusters in the critical
Potts model. This relation is checked numerically by means of cluster
simulations of several O($n$) models in the range $1\leq n \leq 2$.
The simulation results include the percolation threshold for several
values of $n$, as well as the universal exponents associated with
bond dilution and the size distribution of the diluted clusters
at the percolation threshold. 
Our numerical results for the exponents are in agreement with
existing Coulomb gas results for the random-cluster model, which
confirms the relation between both models.
We discuss the renormalization flow of the bond-dilution parameter 
$p$ as a function of $n$, and provide an expression that accurately
describes a line of unstable fixed points as a function of $n$,
corresponding with the percolation threshold. Furthermore, the
renormalization scenario indicates the existence, in $p$ versus $n$
diagram, another line of fixed points at $p=1$, which is stable with
respect to $p$.
\end{abstract}
\maketitle 
\section{Introduction}
\label{intro}
The reduced Hamiltonian of the O($n$) spin model is usually written as
\begin{equation}
\label{Hspin}
{\mathcal H}/(k_{\rm B} T) =
  -J/(k_{\rm B} T) \sum_{<ij>}\vec{S_{i}} \cdot \vec{S_{j}}\, ,
\end{equation}
where $k_{\rm B}$ is the Boltzmann constant, and $T$ the temperature.
The spin $\vec{S_{i}}$ is  an $n$ dimensional vector, localized at
site $i$ of a lattice. The sum is over all nearest-neighbor pairs.
For the special cases $n=1,2,3$, this model corresponds to the Ising, the XY
and the Heisenberg model, respectively.

The O($n$) symmetry actually allows a more general form of the pair
interaction--e.g., one may replace 
$J/(k_{\rm B} T)\vec{S_{i}} \cdot \vec{S_{j}}$
by $a(\vec{S_{i}} \cdot \vec{S_{j}})$ with $a$ an analytic function. 
A particularly useful choice~\cite{loopmodel} is
$a \equiv \ln (1+x \vec{S_{i}} \cdot \vec{S_{j}})$, 
where $x$ is a temperature-like variable.
The so-called high-temperature graph expansion \cite{Onmodel} for
such a model yields a weighted sum of graphs, in which each site
connects to its neighboring sites by an {\em even} number of 
`bonds'. On lattices with coordination number of at most three,
like the honeycomb lattice, these graphs reduce to a set of
non-intersecting loops. For properly normalized O($n$) spins,
the resulting partition sum reads \cite{loopmodel}
\begin{equation}
\label{Zloop}
 Z_{\rm loop}=\sum_{\mathcal G} x^{N_{b}}n^{N_{l}} \, ,
\end{equation}
where the sum is over all possible loop configurations ${\mathcal G}$
that can be constructed on the edges of the lattice.
The number of `bonds', i.e., edges covered by 
${\mathcal G}$, is denoted as $N_b$, and the number of loops in 
${\mathcal G}$ as $N_l$.  Remarkably, the spin dimensionality $n$ of
the original model now appears as a continuous variable. The critical
point of this model is exactly known for $-2 \leq n \leq 2$, as well
as some critical exponents \cite{N1982}.

The significance of clusters in critical or near-critical configurations of
O($n$) and $q$-state Potts models has already been recognized long ago
\cite{droplet, KF,KF1}. While the present work focuses on the percolation
aspects of the two-dimensional O($n$) model, we consider it useful to
first review the similar case of the two-dimensional Potts model.

It is well known that the critical singularities of 
the ferromagnetic $q$-state Potts model 
can be correctly represented in terms of
Kasteleyn-Fortuin (KF) clusters \cite{KF,KF1}, also called random
clusters, rather than in terms of naively defined Potts clusters. 
The latter are formed by connecting, with probability 1, nearest-neighbor
spins in the same Potts state, while the formation of KF clusters uses
a probability $p=1-e^{-K}$ instead, where $K$ is the nearest-neighbor
coupling constant of the Potts model. Thus, one may consider KF clusters
as percolation clusters formed by a bond-percolation process that uses 
the Potts clusters as a substrate.

Both the size of the largest Potts and that of the largest KF cluster
diverge at the critical point, but it is the size of the KF cluster that
determines the spontaneous magnetization, governed by the order-parameter
exponent $\beta(q)$. The Potts magnetic correlation function appears to
be equal to the probability that the correlated sites belong to the same
KF cluster. It follows that the fractal dimension $d_f$ of KF clusters is
equal to the Potts magnetic renormalization exponent $y_h$ \cite{ConKl}.
Potts clusters are denser than KF clusters, and are described by a
larger fractal dimension \cite{CV,DBN}. 
At the Potts critical coupling $K_{\rm c}$, bond dilution of Potts
clusters yields a percolation threshold at bond probability
$p=1-e^{-K_{\rm c}}$ \cite{ConKl,DBN,BKN}, which corresponds 
precisely with KF clusters. In the renormalization language, this
KF point acts as an unstable fixed point on the Potts critical line
parametrized by $p$, and the Potts clusters are described by a
stable fixed point at larger $p$.

For the case of the tricritical $q$-state Potts model, which can be
reached from the pure $q$-state Potts model by including vacant
sites \cite{NBRS,Nientric}, 
the situation is somewhat different. The thermodynamic
singularities are still described by KF clusters, which are obtained
by bond dilution of tricritical Potts clusters with bond probability
$p=1-e^{-K_{\rm t}}$, where $K_{\rm t}$ is the Potts coupling at the
tricritical point. But the percolation threshold of the bond dilution
process no longer coincides with the KF point. The percolation threshold
now occurs at $p<1-e^{-K_{\rm t}}$, and marks an unstable fixed point
with respect to variation of $p$. The fractal properties of tricritical
Potts clusters, as well as those of KF clusters, are now described by a
fixed point that is stable in the $p$ direction.

Furthermore, the pair of fixed points on the tricritical line can
be related to the pair on the critical line. When we define $q'$ as
the number of states for which the tricritical Potts model has the
same conformal anomaly as the $q$-state critical Potts model,
it is found that the universal properties of $q$-state Potts clusters
match those of $q'$-state tricritical KF clusters, and that $q$-state 
KF clusters correspond with diluted tricritical $q'$-state clusters at
the percolation threshold \cite{DBN,JS}.

It is now natural to address a similar percolation problem defined
within the domains separated by critical O($n$) loops. In particular,
one may ask the questions for what bond probability $p$ there
will be a percolation threshold, what is the exponent associated to
$p$, and what is the fractal dimension of the percolation clusters
at the threshold.
While it is possible to find direct answers for the honeycomb O(1)
model, this work will address these questions also for other
values of $n$.
In Sec.~\ref{theory} we predict the exponent $y_h$, which is the
fractal dimension of the clusters at the  percolation threshold,
and $y_p$, or, $\nu_p=1/y_p$, which controls the divergence of the 
percolation correlation length as $\xi = (p-p_{\rm c})^{-1/y_p}$.
Section \ref{algorithm} provides a numerical analysis that yields the
percolation thresholds and a verification of the Coulomb gas result.
The analysis is based on finite-size scaling \cite{FSS} and Monte Carlo
simulations of O($n$) models for several values of $n$, using a recently
developed cluster algorithm \cite{DGGBS}. In Sec.~\ref{selfm} we review
the self-matching lattice argument for the case of the dual triangular
lattice, and its consequences for our percolation problem. We conclude
with a short discussion in Sec.~\ref{concl}.

\section{Bond percolation between dual spins}
\label{theory}

In defining the percolation problem of O($n$) loop configurations, it
is convenient to make use of the representation of such configurations
by means of Ising spins on the dual lattice. Spins not separated by a
loop are given the same sign, and spins separated by one loop have 
opposite signs. Here we ignore the possible inconsistency of 
such an assignment with the existence of periodic boundary conditions,
because we expect that leading critical singularities are not modified
by the restriction that the spins are single valued.
The dual Ising configuration thus completely specifies
the loop configuration (but we do not attempt to express the O($n$)
partition sum in the Ising language).
We can now construct dual clusters by drawing, with probability 1,
bonds between nearest-neighbor sites occupied by spins in the same state.
The hulls of these clusters are the O($n$) loops. At criticality of
the O($n$) loop model, the clusters are fractals with
dimension $d_a$, and the hulls are fractals with dimension $d_l$.

\begin{figure}
\includegraphics[scale=0.3]{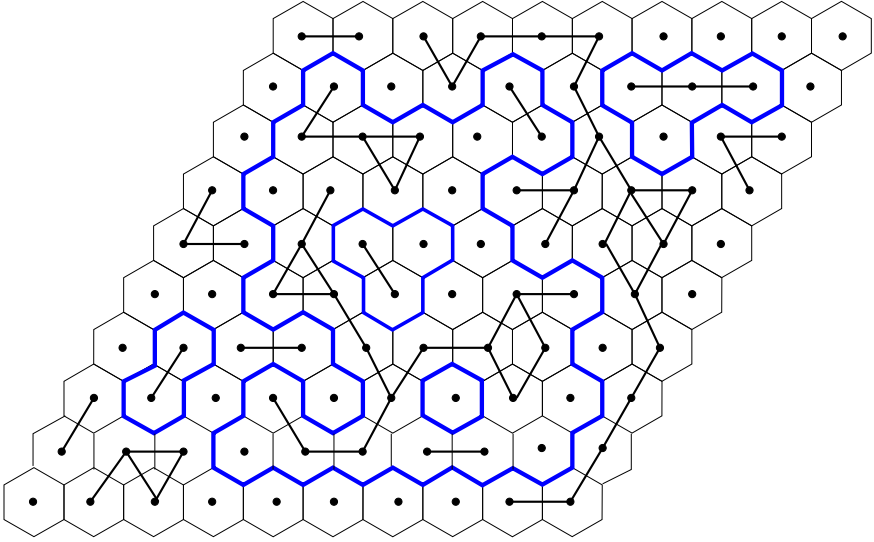}
\caption{An O($n$) loop configuration on the honeycomb lattice,
together with a bond percolation configuration in dual 
clusters as defined within O($n$) loops.}
\label{cfg}
\end{figure}

On the basis of a critical loop configuration, one may now connect
neighboring dual Ising spins in the same state instead with a bond
probability $p<1$ and thus form new types of clusters. This is
illustrated in Fig.~\ref{cfg}.

The results for the percolation problem on Potts clusters described 
in Sec.~\ref{intro} are relevant for the analysis of the geometric
aspects of critical O($n$) loop configurations, because of the well-known
relation between critical O($n$) loops and the hulls of tricritical
random-cluster configurations \cite{N1982,BD}.  As a result,
the magnetic exponent of the tricritical Potts model is equal to the
fractal dimension of the regions separated by loops in the corresponding
O($n$) model on the surrounding lattice.
Although it is not obvious how to interpret the O($n$) loops on the 
honeycomb lattice as the hulls of a KF random-cluster model,
one may assume that the universal aspects of the  percolation problems
within the regions separated by the loops are independent of the 
details of the lattice structure. Moreover, 
the partition sum of the critical O($n$) loop model on the honeycomb lattice
was shown to be identical to that of a tricritical Potts model with
vacancies \cite{Nientric} on the triangular lattice. This mapping
identifies the O($n$) loops with the hulls of tricritical Potts
clusters. As mentioned above, the universal properties of the
tricritical Potts clusters also apply to tricritical KF clusters.
One can thus associate the fractal dimension $d_l$ of critical
O($n$) loops with the hull fractal dimension of  tricritical KF
clusters, and the fractal dimension $d_a$ of the regions separated
by loops with the fractal dimension of tricritical KF clusters.

These exponents are exactly known by means of the Coulomb gas
method \cite{SD,DS,Nienhuis1987}, and verified numerically \cite{DDGQB}
for the O($n$) model.
The critical O($n$) loop model and the equivalent $q=n^2$-state
tricritical Potts model share the Coulomb gas coupling $g$, which  
is given by the following equation
\begin{equation}
\label{cgOn}
q=4 \cos^2(g \pi),~~ 1 \le g \le 2 \, .
\end{equation}
Let $g^{\prime} \equiv 1/g$, thus $0.5 \le g' \le 1$, be the Coulomb
gas coupling of the $q^{\prime}$-state critical Potts model, 
where $q^{\prime}$ is determined by
\begin{equation}
q^{\prime} = 4 \cos^2(g^{\prime}\pi) \, .
\end{equation}
This $q^{\prime}$-state critical Potts model has the same conformal
anomaly as the critical O($n$) loop model, as well as the
$n^2$-state tricritical Potts model \cite{DF,BCN}: 
\begin{equation}
c=1-\frac{6(1-g)^2}{g}=1-\frac{6(1-g^{\prime})^2}{g^{\prime}} \, .
\end{equation}
As already mentioned in  Sec.~\ref{intro}, the Potts clusters of this
$q'$-state critical Potts model are equivalent with the $q$-state KF
clusters of the tricritical Potts model, and therefore also with the
regions separated by critical O($n$) loops.
Just as the process of bond dilution of critical $q'$-state Potts 
clusters leads to KF random clusters, we expect that bond dilution 
of the dual clusters defined within O($n$) loops will, at the bond 
percolation threshold, lead to configurations with critical
KF-like universal properties.
The fractal dimension of the KF clusters of the critical 
$q^{\prime}$-state Potts model is \cite{Nienhuis1987}
\begin{equation}
y_h=1+g'/2+3/(8 g') \, ,
\label{yhp}
\end{equation}
which can be continued analytically into the tricritical
range \cite{BNTcrP}. This holds as well for the bond-dilution exponent,
which is given by \cite{Coniglio1989}
\begin{equation}
y_p=1-3g'/2+1/(2g') \, .
\label{dilution_exp}
\end{equation}

For the special values $n=0$, 1 and 2 it is possible to derive exact
percolation thresholds. 
For $n=0$, loops are in fact forbidden due to their zero weights,
but, depending on the boundary conditions, loop segments may emerge
from the boundaries. In the high-temperature O(0) phase, these
segments will, however, be confined to a boundary layer of finite
thickness, and the bulk of the model will be empty. Therefore,
the percolation threshold is exactly that of the triangular 
bond-percolation model, which is given as the solution of
\begin{equation}
p^3 -3p+1 =0 \, .
\label{n0pc}
\end{equation}
Assuming continuity of the percolation threshold between the 
high-temperature O($n$) phase and the critical state, the bond-percolation
threshold $p_{\rm c}(n=0)$ at O(0) criticality is also equal to the
solution of Eq.~(\ref{n0pc}), which is
\begin{equation}
p_{\rm c}(n=0)= 2 \sin(\pi/18) 
\end{equation}

For the O($1$) loop model, one may apply an exact duality
transformation which yields the critical triangular Ising model. 
Thus, the dual Ising configurations described above are precisely
those of the critical triangular Ising model and we can use its known
properties. The critical point of the triangular model \cite{Hout}
is $K_{\rm c}=(1/4) \ln 3$, and its random-cluster representation
determines the percolation threshold as
\begin{equation}
p_{\rm c}(n=1)=1-\exp(-2K_{\rm c})=1-1/\sqrt{3} \, .
\label{cripoint}
\end{equation}

For the case $n=2$ we apply an argument of a different nature.
First we note that, along the Potts critical line as parametrized by $p$,
the two fixed points, describing the Potts and KF clusters respectively,
merge \cite{DBN} for $q\to 4$, the point where the Potts critical and 
tricritical branches meet. At this point, the difference between the
critical and tricritical KF clusters vanishes. Therefore we expect that
no further bond dilution of the aforementioned dual clusters is
required, i.e., 
\begin{equation}
p_{\rm c}(n=2)=1  \, .
\label{cripoin2}
\end{equation}
The predictions for the percolation thresholds at $n=1$ and 2 
will be the subject of numerical verification in Sec.~\ref{numevcfp}.

\section{Simulation}
\label{algorithm}
\subsection{Sampled variables and finite-size scaling}

The representation of the Potts model by means of KF clusters has led to
the development of cluster Monte Carlo algorithms \cite{SW,W}, which
drastically reduce the critical slowing down problem in simulations of
the Potts model. Since then, more cluster algorithms have been developed,
so that accurate simulation results can now be obtained for a considerable
number of other critical model systems. Here we use an efficient cluster
algorithm \cite{DGGBS} for O($n$) loop models with noninteger $n>1$
to verify the predictions made in Sec.~\ref{theory}. The simulations 
took place at the critical point which is given by
$x_{\rm c}=[2+(2-n)^{1/2}]^{-1/2}$ for the honeycomb lattice \cite{N1982}.
The loop model cluster algorithm \cite{DGGBS} easily allows meaningful
simulations up to a linear system size $L=512$ at the critical point.

The configurations generated by the Monte Carlo algorithm are represented
by means of dual Ising spins. The percolation problem involves the
addition of bonds between equal nearest-neighbor Ising spins with
probability $p$. For this percolation problem on the dual clusters,
we expect, at least in part, a similar
behavior as a function of $p$ as usual in percolation
theory \cite{stauffer}. Thus, for small $p$ the percolation clusters
are small, and they will grow with increasing $p$ until the percolation
threshold $p_{\rm c}$ where the largest percolation cluster diverges,
at least in the thermodynamic limit. In a finite system, the largest
cluster is limited by the system size. When a percolation cluster
reaches the system size, we call it a ``spanning cluster''.
However, this statement has to be made more precise. For a finite
system with periodic boundary conditions, there are different rules
to define a spanning cluster. One may define it as a cluster whose
linear size in at least one of the lattice directions reaches the
size of the periodic box, or as a cluster that connects to itself 
along {\em at least} one of the periodic directions~\cite{Hovi1996}.
Here we use the latter definition. 

In order to obtain the threshold $p_{\rm c}$ and some
critical exponents associated with this percolation problem, 
several quantities  are sampled. These include the susceptibility-like
quantity $\chi_{_G}$, the probability $R_e$ that the occupied bonds form
a nontrivial loop ('nontrivial' means here that the loop spans the 
torus and thus cannot be shrunk into a point by a continuous deformation),
and the density $P_{\infty}$ of the spanning cluster. The quantity $R_e$
is alternatively called the wrapping or the crossing probability.
We provide some further details to describe these quantities.

\begin{figure}[htb]
\includegraphics[scale=0.8]{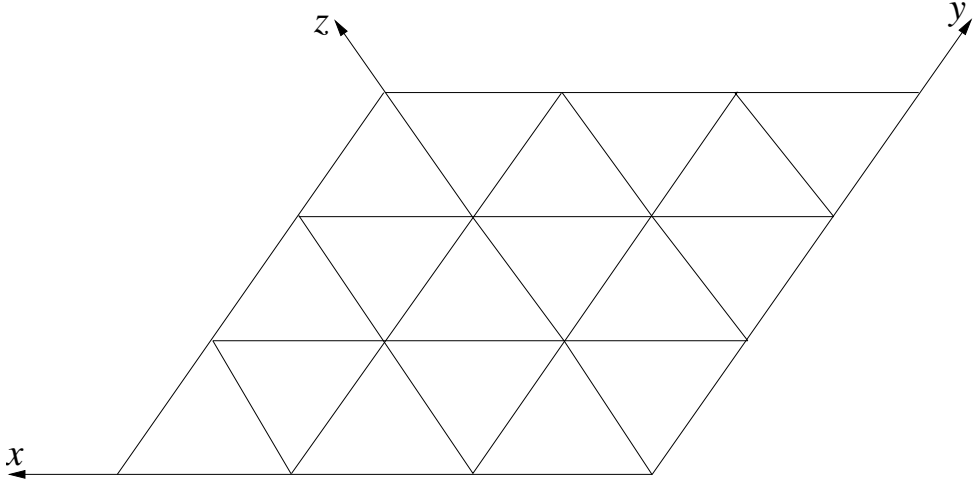}
\centering
\caption{A percolation cluster may span the periodic system with
hexagonal symmetry in the $x$, $y$ and $z$ direction.}
\label{psdef}
\end{figure}
On the triangular lattice, a cluster  may span the system in different
directions, including the ones labeled $x,~y$ and $z$ in Fig.~\ref{psdef}.
We thus define the measure $R_e$ of the spanning probability as 
\begin{equation}
R_e=\langle (R_x+R_y+R_z)/3 \rangle \, ,
\end{equation}
where $\langle \cdots\rangle$ stands for ensemble averaging, 
and the subscript $e$ means that a connection can exist along 
each of the $x$, $y$, and the $z$ directions. 
If there is no cluster that spans the system, we put $R_x=R_y=R_z=0$.
We put $R_x=1$ if there exists of a cluster that
connects to itself over a displacement equal to a unit vector along
the $x$ direction; if there are no other connections to itself in
other directions, we put $R_y=R_z=0$. The same applies with cyclic
permutations of $x$, $y$, and $z$. However, a cluster may also connect
to a periodic image of itself in other directions than those of the
$x$, $y$, and $z$ axes. If the cluster connects to itself with a
nonzero displacement vector that is not parallel to the $x$, $y$, or
$z$ direction, we put $R_x=R_y=R_z=1$.
With this definition of $R_e$, the $x$, $y$, and $z$ directions are
treated equivalently, despite the fact that there are only two
independent directions in two dimensions. 

Finite-size scaling yields the following finite-size behavior for
$R_e$ as a function of the finite size $L$ and the bond probability:
\begin{equation}
 R_e=R_{ec}+a(p-p_{\rm c})L^{y_{_p}}+\cdots+b_1L^{y_{_1}}+    
 b_2L^{y_{_2}}+\cdots \, ,
\label{Psfs}
\end{equation}
where $p_{\rm c}$ is the percolation threshold, 
$y_p$ is the bond-dilution exponent, and $y_1$, $y_2, \cdots$ are
negative correction-to-scaling exponents. The quantity $R_{ec}$ is    
defined as the value of $R_e$ at $p_{\rm c}$, which is also universal
\cite{HTP,ZLK}, but still dependent on the finite system geometry and,
at present, on the underlying O($n$) critical state.

The percolation susceptibility $\chi_{_G}$ and the percolating
cluster density $P_{\infty}$ are defined as 
\begin{eqnarray}
\chi_{_G} &=&L^{-d} \langle \sum^{N_c}_{i}{n^2_i} \rangle,\\
P_{\infty}&=&L^{-d} \langle n_{\infty} \rangle
\end{eqnarray}
respectively,
where $d=2$ represents the dimensionality of the model, and
$N_c$ is the number of clusters.
The number of sites in the $i$-th cluster is denoted $n_i$, and 
$n_{\infty}$ refers to that in the spanning cluster. 
Thus $P_{\infty}$ 
represents the probability that a randomly chosen site belongs to
the spanning cluster. The definition of $\chi_{_G}$ is the same
as the Potts magnetic susceptibility, expressed in terms of 
an ensemble average over random clusters.

Finite-size scaling  predicts the following behavior for
 $P_{\infty}$ and $\chi_{_G}$:
\begin{eqnarray}
P_{\infty}&=&L^{y_{_h}-d}(a_0+a_1(p-p_{\rm c})L^{y_{_p}}+\cdots+
b_1L^{y_{_1}}+b_2L^{y_{_2}}+\cdots) \, ,\label{pfs0}\\
\chi_{_G}&=&L^{2y_{_h}-d}(a_0+a_1(p-p_{\rm c})L^{y_{_p}}+\cdots+
b_1L^{y_{_1}}+b_2L^{y_{_2}}+\cdots) \, ,\label{chifs0}
\end{eqnarray}
where $y_h$ is the fractal dimension of the percolating cluster. 
At the percolation threshold $p_{\rm c}$, these equations reduce to 
\begin{equation}
P_{\infty}= L^{y_{_h}-d}(a_0+b_1L^{y_{_1}}+b_2L^{y_{_2}}+\cdots)\, ,
\label{pfs}\\
\end{equation}
and
\begin{equation}
\chi_{_G}= L^{2y_{_h}-d}(a_0+b_1L^{y_{_1}}+b_2L^{y_{_2}}+\cdots)\, .
\label{chifs}
\end{equation}

\subsection{Results}
\label{resul}
We illustrate the numerical procedure, using the O($1.5$) loop model as
an example. The model with periodic boundary conditions was simulated at
its critical point \cite{N1982}, which is $x_{\rm c}=0.60778\cdots$.
Since the cluster algorithm hardly suffers from any critical slowing
down, as described in Ref.~\cite{DGGBS}, samples were taken at intervals
of only $2$ cluster steps.

\begin{figure}[htbp]
\includegraphics[scale=1.6]{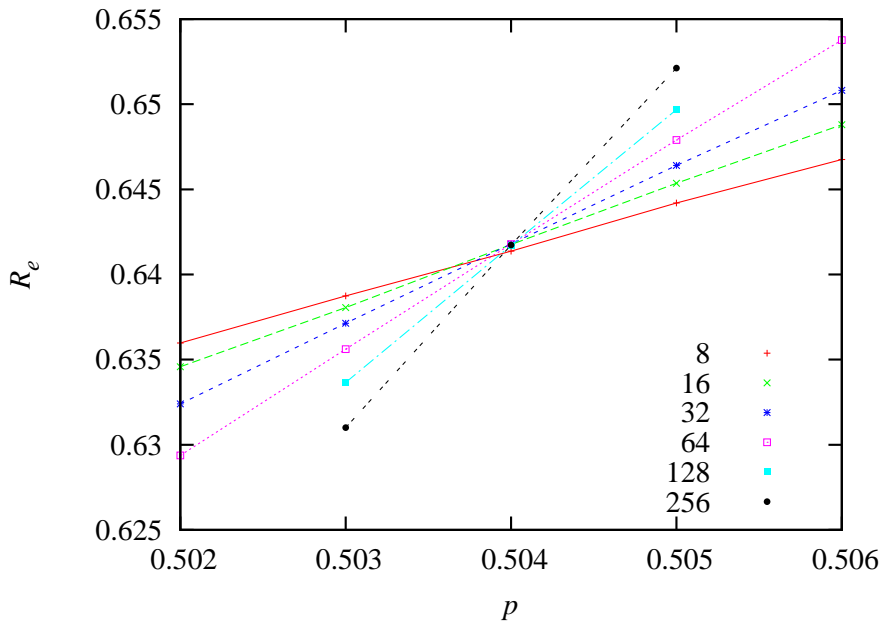}
\caption{ Spanning probability $R_e$ versus bond probability $p$
for various system sizes in the case $n=1.5$. The lines connecting the
data points are added only for illustration. All error bars are much
smaller than the size of the data points. }
\label{ps15}
\end{figure}

The first stage involved the determination of the spanning probability
$R_e(p,L)$, the density $P_{\infty}$ of the percolating cluster,
and the percolation susceptibility $\chi_{G}$ 
for several values of the bond probability $p$. This was
done for $6$ system sizes $L$ ranging from $8$ to $256$. 
After the equilibration of the system, $10^{8}$ samples were taken
for each value of $p$ in the range $8\leq L\leq 64$, and $4\times 10^{7}$
samples in the range $64<L\leq 256$.
Statistical errors were estimated by dividing each run in $1000$
partial results, and subsequent statistical analysis. 
The correlations between subsequent partial results are negligible for
the lengths of these runs.
Parts of the $R_e$ data are shown in Fig.~\ref{ps15}. 

The number of free parameters in the finite-size scaling equations
(\ref{Psfs}), (\ref{pfs0}) and (\ref{chifs0}) makes it necessary to apply a
multivariate method. We used the Levenberg-Marquardt least-squares
algorithm, which allows nonlinear fits according to these 
equations. We thus determined the percolation threshold $p_{\rm c}$
from the $R_e$ versus $p$ data.
The next step involved a simulation of seven system sizes in the
range $8\leq L\leq 512 $ at our estimated value of $p_{\rm c}$,
with the same lengths as mentioned above for $8\leq L\leq 256$, and
$10^{7}$ samples for $L=512$. Including these runs, we fitted the
unknowns in the finite-size scaling formula Eq.~(\ref{Psfs}), 
also including the universal probability $R_{ec}$ and the dilution 
exponent $y_p$ to the data. This yielded our final estimates, 
namely $p_{\rm c}=0.50403~(1)$, $R_{ec}=0.6420(2)$, and
$y_p=0.395~(2)$ for the bond-dilution exponent.
The latter result is in a good agreement
with the Coulomb gas prediction $y_p=0.3955\cdots$.

\begin{figure}[htbp]
\includegraphics[scale=1.6]{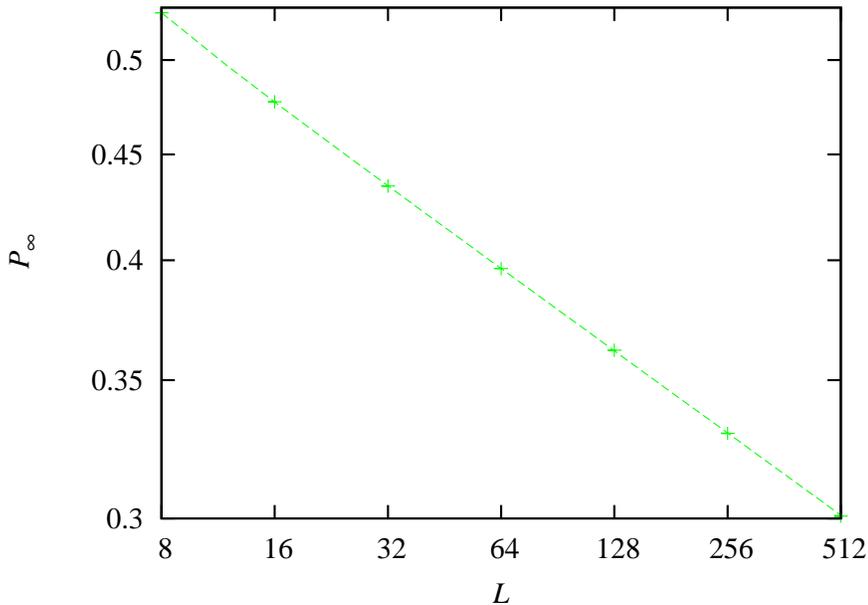}
\caption{Density $P_\infty$ of the spanning cluster at the percolation
threshold versus system size
$L$ for the case $n=1.5$, shown on logarithmic scales. The dashed line
represents a fit to the data points according to Eq.~(\ref{pfs}).
The error bars are much smaller than the size of the data points.  }
\label{p15}
\end{figure}

Since the analysis of the data for the density $P_{\infty}$ of the 
percolating cluster and for the percolation susceptibility $\chi_{G}$
yielded the $p$ dependence in terms of the coefficients $a_i$, we
can deduce the value of $P_{\infty}$ and  $\chi_{G}$ at our best
estimate for the percolation threshold.
Parts of these data for $P_{\infty}$ are shown in Fig.~\ref{p15}
versus $L$.
We find that, at the percolation threshold, $P_{\infty}$ and $\chi_{_G}$
are well described by power
laws as a function of sufficiently large lattice sizes $L$, in
agreement with the finite-size scaling behavior expressed by
Eqs.~(\ref{pfs}) and (\ref{chifs}). 
A fit of the numerical data for $P_\infty$ and $\chi_{_G}$
according to these equations
yields the fractal dimension $y_h$ of the percolation clusters as
$y_h=1.8678~(2)$ and $y_h=1.8678~(1)$ respectively.
Both values are in a good agreement with the value $1.86775\cdots$ 
based on the Coulomb gas prediction, Eq.~(\ref{yhp}).

The same procedure as described above was applied to the cases 
$n=1.0$, $1.25$, $\sqrt{2}$, $1.5$, $1.75$,$1.90$ and $1.95$.
The percolation thresholds $p_{\rm c}$,
the bond-dilution percolation exponent $y_p$, and the fractal 
dimensions of the percolation cluster $y_h$ are obtained similarly.
The results are listed in Table \ref{finaltab}. The exponents are in a
satisfactory agreement with the values predicted by Eqs.~(\ref{yhp})
and (\ref{dilution_exp}).

\begin{figure}
\includegraphics[scale=1.6]{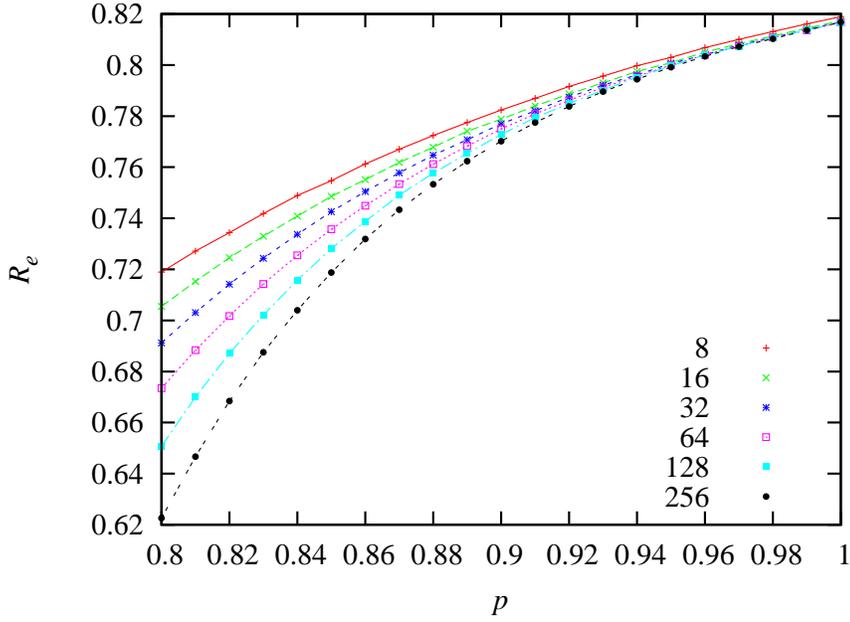}
\caption{
Spanning probability $R_e$ versus bond probability $p$
for various system sizes in the case $n=2$. The lines connecting the
data points are added only for illustration.}
\label{psn2}
\end{figure}

We also performed simulations for $n=2$. However,
the $R_e$ versus $p$ data, which are shown in Fig.~\ref{psn2},
do not show intersections for different system sizes.
Near $p=1$, the curves appear to run asymptotically parallel at vanishing 
distances for large $L$. These findings are consistent with the predicted
values $p_{\rm c}=1$ and $y_p=0$. However, the vanishing of $y_p$ renders
Eq.~(\ref{Psfs}) insufficient for a numerical determination of $p_{\rm c}$
from the $R_e$ data for $n=2$.  We thus fitted the exponent $y_h$ using the
data for $P_\infty$ and $\chi_{_G}$ at the theoretical value $p_{\rm c}=1$.
These results are included in Table \ref{finaltab}. Before attempting a
numerical analysis of the $R_e$ data, we will adapt Eq.~(\ref{Psfs}) for
the case of a marginally relevant bond dilution field, with the help
of renormalization considerations in Sec.~\ref{selfm}.

\section{Self-matching argument and renormalization flow}
\label{selfm}
\subsection{Matching property}
First we shall briefly review the matching properties 
\cite{Sykes-64,Es} of planar lattices and the consequences for the
site-percolation thresholds of some of these lattices.
Let ${\mathcal P}\equiv ({\mathcal V}, {\mathcal A})$ be a planar
lattice, where ${\mathcal V}$ is the set of lattice sites and
${\mathcal A}$ the set of edges connecting the nearest-neighbor sites
of ${\mathcal V}$. The faces of this lattice are polygons without any
`diagonals'. Let ${\mathcal B}$ be the set of diagonals connecting all
pairs of non-nearest-neighboring sites within each polygon. Then, we
define the lattice ${\mathcal L}$, in which these diagonals are included,
as ${\mathcal L}\equiv ({\mathcal V}, {\mathcal A}+{\mathcal B})$.
Then, ${\mathcal P}$ and ${\mathcal L}$ are called matching lattices.
It is possible to define matching lattices in a more general way, but
that is unnecessary for our present purposes.

Suppose that a given site-percolation configuration on ${\mathcal P}$
percolates in the $\vec{x}$ lattice direction. The existence of a
percolating path prevents the existence of a percolating path in a
conjugate percolation configuration obtained as follows.
Replace the occupied sites of ${\mathcal V}$ by empty sites, and
vice versa. Add the diagonals, leading to the lattice ${\mathcal L}$,
and consider the percolation problem in the other lattice direction
denoted $\vec{y}$. It then follows that there is no percolating path
in that direction. Furthermore, if a given configuration of site
variables on ${\mathcal P}$ does not lead to a percolating path in
the $\vec{x}$ direction, then the conjugate problem must have a
percolating path in the $\vec{y}$ direction. As a consequence,
if the site percolation threshold of lattice ${\mathcal P}$ is 
$p^{\rm (s)}_{\mathcal P}$, and the site percolation threshold
of the lattice ${\mathcal L}$ is $p^{\rm (s)}_{\mathcal L}$, then
the thresholds are related as
\begin{equation}
p^{\rm (s)}_{\mathcal P}+p^{\rm (s)}_{\mathcal L}=1 \, .
\label{pml}
\end{equation}

Since no `diagonals' can be added into the triangular lattice, the
triangular lattice is called a self-matching lattice, and the
difference between the two thresholds in Eq.~(\ref{pml}) vanishes.
The matching argument thus yields that the percolation threshold 
of the triangular lattice, as well as that of other self-matching
lattices, lies at $p^{\rm (s)}_c=1/2$.

An important feature of the matching argument is its independence of
interactions between the site variables, as long as these interactions
are symmetric under the interchange of occupied and unoccupied lattice
sites.

\subsection{Percolation at bond probability $p=1$}
\label{perp1}
For the case $n=1$ and bond probability $p=1$, the model reduces to the
Ising model on the triangular lattice with coupling $K = -(\ln x)/2 $, and
all neighboring sites with equal spins are connected by occupied bonds.
The $x <x_c=1/\sqrt{3}$ region corresponds to the low-temperature 
ferromagnet, and $1 \geq x \geq x_c$ to the high-temperature ferromagnet,
and $x >1$ to the antiferromagnet.
Thus, the symmetry between $+$ and $-$ spins holds in the range
$x \geq x_c$. If one considers $+$ spins
as occupied sites, and $-$ spins as unoccupied ones, the bond percolation 
model with $p=1$ can be regarded as a correlated site percolation model
with site-occupation probability $p^{\rm (s)}=1/2$. 
In particular, the $x=1$, i.e., $K=0$ case reduces to the standard site
percolation on the triangular lattice \cite{SD}.
The aforementioned self-matching relation tells us that the whole line 
for $p=1$ and $x \geq x_c$ is a critical line of the percolation type,
as already noted in Ref.~\cite{Qian-2005}. Figure 5 in the latter
reference describes
the renormalization flow of the model in the $p$ versus $K$ plane.
Since the percolation critical line must be a flow line, 
it follows that, at the Ising critical temperature, the
point $p=1$ is a fixed point. Since the KF fixed point is unstable, 
and there is no sign of intermediate fixed points \cite{Qian-2005},
the $p=1$ fixed point is stable along the $p$ direction. 

For $n \neq 1$, the O($n$) loop model involves non-local interactions,
as reflected by the quantity $N_{l}$ in Eq.~(\ref{Zloop}).
Nevertheless, the symmetry between the $+$ and $-$ spins in the dual
triangular lattice still holds as long as $x \geq x_c (n)$.
The matching argument yields that, for $ n \leq 2$, the point
$p=1, x=x_c (n)$ is always a fixed point for the flow along the
$p$ direction.
 
Consider the subspace $(n,p)$ with $x=x_c (n)$ in the three-parameter 
space $(n,p,x)$. We have now derived two lines of fixed points as 
a function of $n$, namely $p=1$ from the self-matching argument, and 
$p=p_c (n)$ from our numerical estimates. The latter seems to be
tangent to the line $n=2$, 
while the former is perpendicular to the $n=2$ line. This tells
something about the renormalization flow near $n=2$.

\begin{figure}[htb]
\includegraphics{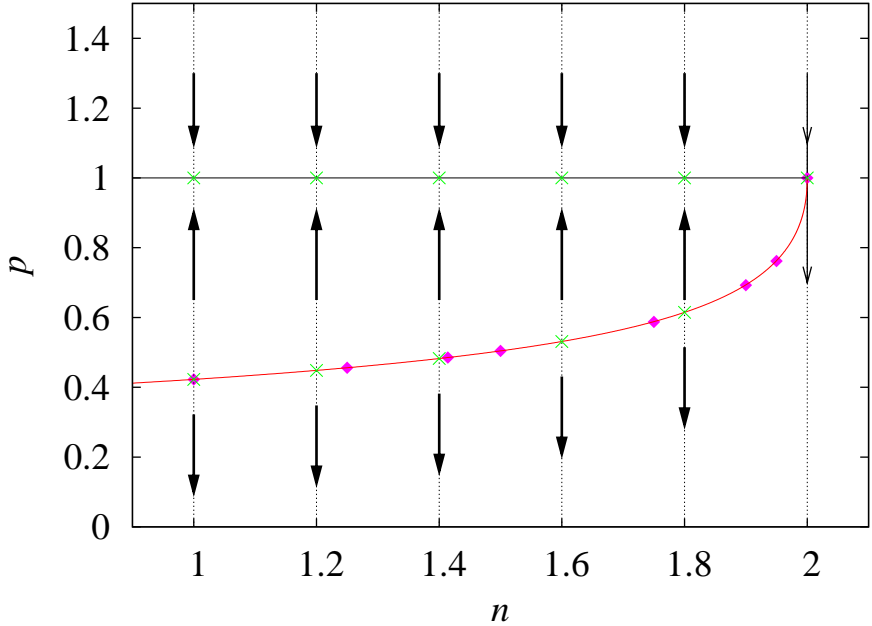}
\centering
\caption{Percolation threshold $p_c (n)$ and the conjectured renormalization
flow. The numerical estimates of $p_c (n)$ are shown as diamonds,
and the $\times$ symbols stand for the fixed points.
The renormalization flow is represented by the arrows, and for $n=2$ we
use the thin arrows for the marginally relevant field.
The curve is given by Eq.~(\ref{fit_pc}) with $a_0 =0.486, a_1=1.5$. }
\label{pc}
\end{figure}

According to the aforementioned results, we conjecture the associated 
renormalization flow as shown in Fig.~\ref{pc}. For $n=2$, the bond-dilution 
field is marginally relevant for $p <1$ and marginally irrelevant $p>1$. 
The lowest order renormalization equation in $\Delta p\equiv 1-p$ and
$\sqrt{\Delta n} \equiv \sqrt{2-n}$ leading to the flow diagram sketched in
Fig.~\ref{pc} is
\begin{equation}
\frac{d \Delta p}{d l}=r_0 \Delta p \sqrt{\Delta n} + r_1
(\Delta p)^2
\label{rgflow}
\end{equation}
where $l$ parametrizes the renormalization flow such that the
rescaling factor is $e^l$, and $r_0$ and $r_1$ are unknown constants.
The appearance of $\sqrt{\Delta n}$ in this equation is in line with
the dependence of the O($n$) critical point $x_{\rm c}$ and that of
the Coulomb gas coupling constant $g$ on $n$.

\subsection{Numerical evidence for the conjectured fixed points
at $p (n)=1$}
\label{numevcfp}
If the line of stable fixed points is indeed located at
$p (n)=1, x=x_c(n)$ for $ 0 \leq n \leq 2$,
the amplitude of the irrelevant bond-dilution field
is zero at and only at $p=1$.
For a test, we simulated the critical $n=1.5$ loop model near $p=1$.
The bond-dilution exponent is given by Eq.~(\ref{dilution_exp}),
where the Coulomb-gas couplings for the stable and unstable
fixed points relate as $g g' =1$. For $n=1.5$ this yields
the bond-dilution exponent as $y_p'=-0.4386$ near $p=1$ and 
$y_p=0.3955$ near the threshold $p_c$.

\begin{figure}[htb]
\includegraphics[scale=0.9]{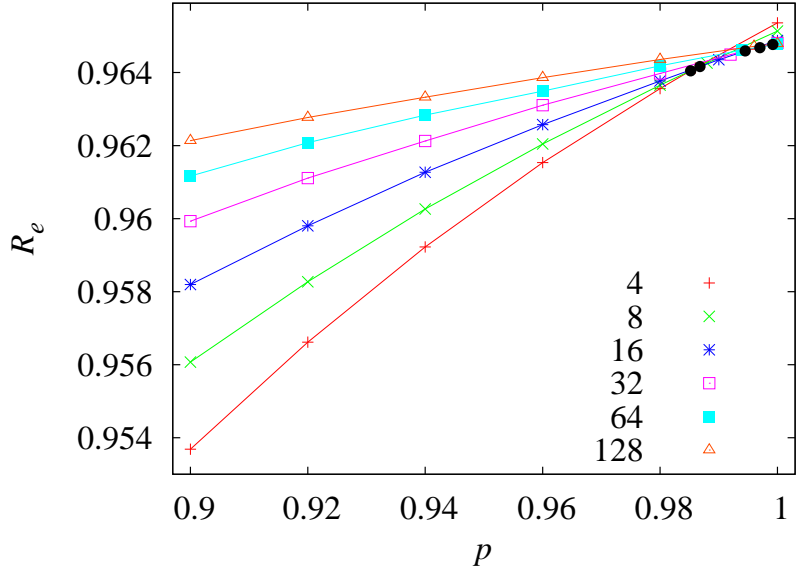}
\centering
\caption{Spanning probability $R_e$ for $n=1.5$ and near $p=1$ vs.
the bond-occupation probability $p$. The labels represent linear size
$L$. The filled circles approximately locate at intersections of the
$L$ and the $2L$ data lines with $L=4,8,16,32,64$. They rapidly approach
to $p=1$.}
\label{p1n1-5-0}
\end{figure}

Parts of the $R_e$ data near $p=1$ are shown in Fig.~\ref{p1n1-5-0}
versus the bond-occupation probability $p$.
The $R_e$ data lines become more and more flat when size $L$
increases, reflecting that the bond-dilution field is
irrelevant. The filled circles mark
the intersections of the $L$ and the $2L$ data lines with
$L=4$, 8, 16, 32 and 64. They are approaching $p=1$ when $L$ increases,
suggesting that $p=1$ is indeed a stable fixed point.
To obtain more solid evidence, we fitted the
$R_e$ data at $p=1$ by $R_e (L) =R_{e1}+b L^{y_c}$, and obtained
$R_{e1}=0.964780~(8)$ and $y_c =-1.9~(2)$, which indicates the absence
of finite-size corrections with exponent $y_p'=-0.4386$.

\begin{figure}[htb]
\includegraphics{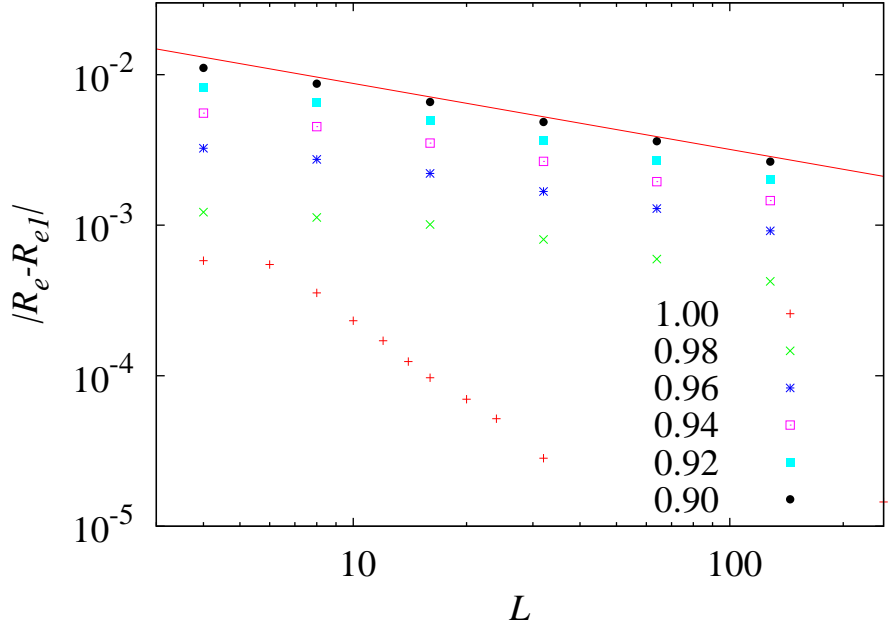}
\centering
\caption{Spanning probability, shown as $|R_e-R_{e1}|$ for $n=1.5$
vs. system size $L$, using logarithmic scales, for several values of
the bond-occupation probability $p$. Different symbols correspond
with different values of $p$ as listed in the figure.
The value $R_{e1}=0.964780$ was taken from the least-squares fit.
For comparison, we show a straight line with a slope $y_p'=-0.4386$,
describing the finite-size dependence due to a correction to scaling
associated with the irrelevant field along the $p$ direction.}
\label{p1n1-5-1}
\end{figure}

The data are also shown in Fig.~\ref{p1n1-5-1} as $|R_e (L,p)-R_{e1}|$
versus $L$, where $R_{e1}=0.964780 $ was taken from the fit.
This figure illustrates that
the difference $|R_e (L,p)-R_{e1}|$ at $p=1$ vanishes much
more rapidly than that for $p \neq 1$. It also rather clearly
demonstrates that the finite-size correction exponent is independent
of the  bond-occupation probability $p$ when $p \neq 1$, and that its
value is in agreement with the expected value $y_p'=-0.4386$.

We observed that the $R_e$ data in range $10 \leq L \leq 512$
can be well described by
\begin{equation}
R_e(p,L) =R_{e1} + a_1 (p-p_{c1}) L^{y_p'} + a_2 (p-p_{c1})^2
L^{2y_p'} + b L^{-2} +c L^{-2+y_p'} \; .
\label{fit_re_p1}
\end{equation}
The fit yields $y_p'=-0.40~(3)$, $R_{e1}=0.96477~(3)$
and $p_{c1}=0.9996~(6)$,
consistent with the values found from the fit of the data at $p=1$,
and in good agreement with with the expected value $p_{c1}=1$.

Furthermore, we investigate whether the numerical data for $n=2$ are
consistent with the existence of a marginal fixed point at $p=1$.
Integrating the renormalization flow, Eq.~(\ref{rgflow}), for $n=2$,
and setting the finite size $L$ equal to the rescaling factor $e^l$,
one finds the renormalized value $\Delta p'$ as
\begin{equation}
\Delta p'= \frac{\Delta p}{1-r_1 \Delta p \ln L} \; .
\label{delpp}
\end{equation}
The corresponding finite-size-scaling equation for the spanning
probability is $R_e(\Delta p,L,u)=R_e(\Delta p',1,L^{y_i}u)$,
where $u$ is an irrelevant field and $y_i$ its renormalization 
exponent. Substitution of $\Delta p'$, and expansion of the scaling
function in small arguments, yields
\begin{equation}
R_e(\Delta p,L,u)= R_{ec} +R_1 \Delta p' + R_2 (\Delta p')^2 + \cdots +
b_1 L^{y_i}
\label{Rsc}
\end{equation}
The numerical data for $R_e$ were fitted by this formula, and several
variations of it, concerning the number of irrelevant fields, and the
degree in $\Delta p'$. The data for small $\Delta p$ reveal the existence
of only one irrelevant field, with an exponent $y_i$ close to $-2$. 
With this exponent fixed at $y_i=-2$, we obtain fits with satisfactory
residuals for the data with $R_e\geq 0.74$. The fits deteriorate
for cutoffs at smaller values, apparently because the expansion
parameter $\Delta p'$ in Eq.~(\ref{Rsc}) becomes too large.
We find that $R_{ec}=R_{e1}=0.8168~(5)$
at the marginal fixed point for $n=2$, and that it lies at
$\Delta p_c\equiv 1-p_c =-0.001 \pm 0.002$, in agreement with the
expected location $p_c=1$.
These data are included in Table \ref{finaltab}.

\subsection{Numerical representation of the percolation threshold}
For a description of the numerical estimates of the percolation
thresholds as a function of $n$, we impose three conditions: \\
(1) $\Delta n \propto (\Delta p)^2$ for $\Delta n \rightarrow 0$;\\
(2) $\Delta p =1/\sqrt{3}$ for $\Delta n=1$;\\
(3) $\Delta p = 1-2 \sin (\pi /18)$ for $n \rightarrow 0$.\\
These conditions are based on the exact values given in
Sec.~\ref{theory}, which are supported and, for condition (1),
supplemented by our numerical evidence.
The second condition is a solution of
$f_1 (\Delta p ) \equiv 1 - 2 (\Delta p )^2 -3  (\Delta p )^4 =0$, 
and the third one of
$f_0 (\Delta p ) \equiv 1 - 3 (\Delta p )^2 + (\Delta p )^3=0$.

\begin{figure}[htb]
\includegraphics{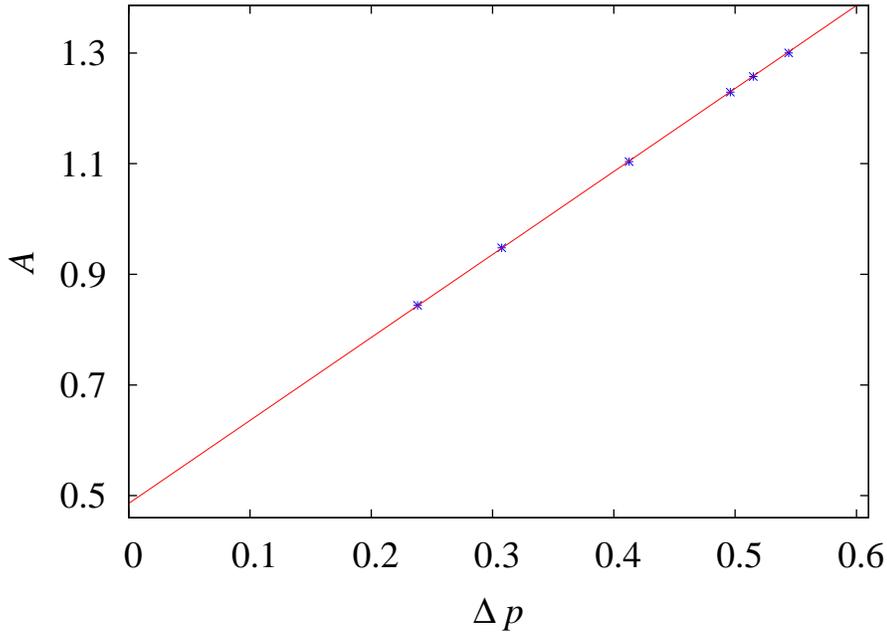}
\centering
\caption{Amplitude $A$ in Eq.~(\ref{fit_pc}).
The data points are for $n=1.25$, $\sqrt{2}$, $1.5$, $1.75$,
$1.9$, and $1.95$ when $\Delta p$ decreases.
The straight line is drawn as $A = a_0 +a_1 \Delta p$, with
$a_0=0.488$ and $a_1=1.493$.}
\label{pc1}
\end{figure}

We fitted the $p_c(n)$ data in Table I by the formula 
\begin{equation}
\Delta n= \frac{f (\Delta p)} { f_1 (\Delta p) + f (\Delta p)} \; ,
\label{fit_pc}
\end{equation}
with $f (\Delta p) \equiv (\Delta p)^2 \left[ 2 (\Delta p)^3 
\left(4- \Delta p \right) + f_0 (\Delta p) A ( \Delta p ) \right]$, 
where the amplitude $A$ is a polynomial of $\Delta p$--namely,
$A(\Delta p ) = a_0 +a_1 \Delta p +a_2 ( \Delta p )^2 + \cdots $.
This equation exactly reproduces 
$f_1=0$ for $\Delta n=1$ and $f_0 =0$ for $\Delta n=2$. 
From our numerical data $p_c (n)$, we calculated the 
amplitude $A ( \Delta p )$, shown in Fig.~\ref{pc1}. The fit 
for the $A$ data yields $a_0=0.488$, $a_1=1.493$, and $a_2 =0.000$.
Although $ a_0$ and $a_1$ lie quite close to the simple fractions
$1/2$ and $3/2$, it is very unlikely that they are equal to these
values, according to the $\chi^2$ criterion.

We conclude this section with a comment on the condition (1). 
For $\Delta n>0$, Eq.~(\ref{rgflow}) yields two fixed points, namely
$(\Delta p)_0=0$ and $ -\sqrt{\Delta n}r_0/r_1$. Expansion of
Eq.~(\ref{rgflow}) near these two fixed points gives the bond-dilution 
exponents $y_p' = r_0 \sqrt{\Delta n} $ near $(\Delta p)_0=0$
and $y_p = - r_0 \sqrt{\Delta n} $ for $(\Delta p)_0=
-r_0\sqrt{\Delta n} /r_1$. Using the relation between $g$ and $n$ and
Taylor-expansion of Eq.~(\ref{dilution_exp}) near $n=2$,
one obtains $y_p' = - 2 \sqrt{\Delta n} / \pi$ near $(\Delta p)_0=0$
and $y_p  =  2 \sqrt{\Delta n} / \pi$ near $(\Delta p)_0=
-r_0\sqrt{\Delta n} /r_1$. Therefore, one has $r_0=-2/\pi$.
The amplitudes $r_0$ and $r_1$ relate to
$a_0$ in Eq.~(\ref{fit_pc}) as $a_0 = (r_1)^2/(r_0)^2$.

\begin{table}[htbp]
\caption{Numerical results (N) for the percolation threshold $p_{\rm c}$, 
the bond dilution exponent $y_p$, and the fractal dimension $y_h$  
of the percolation problem on critical O($n$) loop configurations.
In the absence of intersections as in Fig.~\ref{ps15}, the entry for
$p_{\rm c}$ at $n=2$ was roughly estimated from Fig.~\ref{psn2}.
Theoretical predictions(T) are included where available.}
   \begin{tabular}{c |l| l l l l l }
    \hline
    \hline
      $n$& & $p_{\rm c}$     &$ y_p$    &$R_{ec}   $ &$y_{_h}$(from $P_{\infty}$) 
& $y_{_h} $ (from $\chi_{_G}$) \\
\hline 
    1.00 &N~~ &0.42265~(1)~&0.542~(2)~ &0.5660~(2)~&1.8749~(2)~&1.8750~(1) \\
         &T~~ &0.422649    &0.5416     &-          &15/8       &15/8       \\
\hline
    1.25 &N~~ &0.45587~(2) &0.475~(1)  &0.6018~(2) &1.8709~(3) &1.8710~(1) \\
         &T~~ &-           &0.4753     &-          &1.87098    &1.87098    \\
\hline
$\sqrt{2}$&N~~&0.48508~(2) &0.425~(1)  &0.6275~(2) &1.8687~(3) &1.8687~(1) \\
         &T~~ &-           &0.4250     &-          &1.86875    &1.86875    \\
\hline
    1.50 &N~~ &0.50403~(1) &0.395~(2)  &0.6420~(2) &1.8678~(2) &1.8678~(1) \\
         &T~~ &-           &0.3955     &-          &1.86775    &1.86775    \\
\hline
    1.75 &N~~ &0.58745~(4) &0.290~(3)  &0.6932~(2) &1.8658~(3) &1.8660~(1) \\
         &T~~ &-           &0.2882     &-          &1.86603    &1.86603    \\
\hline
    1.90 &N~~ &0.6924~(2)  &0.186~(5)  &0.7402~(3) &1.8668~(4) &1.8670~(1) \\
         &T~~ &-           &0.1882     &-          &1.86700    &1.86700    \\
\hline
    1.95 &N~~ &0.7621~(5)  &0.13~(1)   &0.7650~(5) &1.868~(1)  &1.8684~(2) \\
         &T~~ &-           &0.1355     &-          &1.86845    &1.86845    \\
\hline
    2.00 &N~~ &1.001~(2)   &$-0.01$~(2)&0.817~(1)  &1.8750~(3) &1.87500~(1)\\
         &T~~ &1           &0          &-          &15/8       &15/8       \\
\hline
\end{tabular} 
\label{finaltab}
\end{table}

\section{Conclusion and Discussion}
\label{concl}
According to the evidence presented in Sec.~\ref{theory}, the dual
clusters defined in the critical O($n$) loop model should have the
universal properties of KF and Potts clusters in the tricritical
Potts model, under the condition that the O($n$) loop model and the
tricritical $q$-state  Potts model have the same conformal anomaly,
which implies that $q=n^2$. Thus we also deduced that dilution of the
dual clusters in the O($n$) loop model by means of a bond-percolation
process leads to a percolation transition
with the same universal properties as the `geometric' fixed point of
diluted KF clusters in the tricritical Potts model \cite{DBN}.
Moreover, these universal properties should also be the same as KF
clusters in the critical $q$-state Potts model, again under the
condition that the conformal anomaly is the same.
The subsequent numerical verification in Sec.~\ref{algorithm}
confirms these predictions in satisfactory detail. 
The diagram below shows the universal relations between the various
systems by means of vertical arrows, and the effect of dilution
is indicated by horizontal arrows.
\begin{eqnarray}
{\rm O}(n)~{\rm dual~clusters~}~~~~~             &\rightarrow&    ~~~~
{\rm dilute~O}(n)~{\rm dual~clusters}                             \nonumber\\
\updownarrow ~~~~~~~~~~~~~~~& &~~~~~~~~~~~~~~~   \updownarrow     \nonumber\\
{\rm tricritical~KF~clusters}~~~~                &\rightarrow&    ~~~~
{\rm  dilute~tricritical~clusters}                                \nonumber\\
\updownarrow ~~~~~~~~~~~~~~~& &~~~~~~~~~~~~~~~   \updownarrow     \nonumber\\
{\rm Potts~critical~clusters}~~~~                &\rightarrow&    ~~~~~~~ 
{\rm critical~KF~clusters}                                        \nonumber
\end{eqnarray}

\ack
We are much indebted to Bernard Nienhuis for valuable discussions.
This research is supported by the National Science Foundation of
China  under Grant \#10675021, by the Science Foundation of The Chinese 
Academy of Sciences, and by the Beijing Normal University
through a grant as well as support from its HSCC (High Performance
Scientific Computing Center). YD also thanks the support by
the Program for New Century Excellent Talents in University (NCET).

\end{document}